\documentclass[aps,amsmath,amssymb,nofootinbib,12pt]{revtex4}
\usepackage{graphicx,dcolumn,booktabs}
\usepackage{amsmath}
\usepackage{amsfonts}
\usepackage{amssymb}
\usepackage{times}
\usepackage{amsmath, amssymb,color}

\begin{document}
%\begin{CJK}{GBK}{}
\title{Study on the mixing among the $0^{++}$ mesons around $1\sim 2$ $\mathrm{GeV}$ with the QCD sum rules}

\author{ Xu-Hao Yuan  \footnote{segoat@mail.nankai.edu.cn},
         Liang Tang \footnote{tangliang@mail.nankai.edu.cn}
         Mao-Zhi Yang \footnote{yangmz@nankai.edu.cn}
         Xue-Qian Li  \footnote{lixq@nankai.edu.cn} }

\affiliation{
  School of Physics, Nankai University, Tianjin 300071,
  China }

\begin{abstract}
\noindent We calculate the correlation functions of $0^{++}$
$q\bar q$, $s\bar s$ and glueball in the QCD sum rules and obtain
the mass matrix where non-diagonal terms are determined by the cross
correlations among the three states. Diagonalizing the mass matrix and identifying
the eigenstates as the physical $0^{++}$ scalar mesons,
we can determine the mixing. Concretely, our calculations determine the fractions of $q\bar
q$, $s\bar s$ and glueball in the physical states
$f_0(1370)$, $f_0(1500)$ and $f_0(1710)$, the results are  consistent
with that gained by the phenomenological research.
\end{abstract}
%\pacs{12.39.Pn,14.40.Pq,11.10.St}

\maketitle

\section{Introduction}

Existence of glueball is a long-standing puzzle in the QCD theory.
Searching for it becomes the most challengeable task for high energy
physics society. The QCD theory predicts its existence and the
lattice QCD almost determines the mass spectra of glueballs with
various quantum numbers\cite{Ishikawa:1982tb,Berg:1982kp,de
Forcrand:1984rs,Teper:1987wt,Albanese:1987ds,Teper:1998kw,Morningstar:1999rf,Ishii:2002ww,Meyer:2004jc}.
It is believed that the mass of the lighter glueballs should be at
around 1 to 2 GeV. But where are they, can we pin down them? Several
bound states near 2 GeV  have been found in recent
experiments\cite{Nakamura:2010zzi}.  People believe that the number of these
states indeed exceeds that predicted by the simple symmetry
analysis. One natural explanation is that there exist exotic states
and the newly observed resonances are either such exotic states,
glueball, hybrid and multi-quark states, or their mixtures. In fact,
none of the resonances which are newly observed at BES and BELLE can
be identified as glueballs, so that one is tempted to conclude that
glueballs mix with the regular quark states. The lattice and other
model-dependent calculations all predict the mass of the $0^{++}$
glueball falling within the range of about
$1.7\mathrm{GeV}$\cite{Bali:1993fb,Yuan:2009vs,Bagan:1990sy}.
Meanwhile the mass of the state made of pure light quarks $\bar qq$,
where, the $q$ refers to  u, d and s quark, is also near
$1.3\sim1.7\mathrm{GeV}$\cite{Reinders:1981ww,Burcham:1995jg,ROSENFELD:1967zz},
therefore it is very possible that the scalar glueball and the quark
states mix to constitute physical states. The observed resonances
$f_0(1370)$, $f_0(1500)$ and $f_0(1710)$ which have masses close to
$[1.3\sim1.7]\mathrm{GeV}$, can really be such mixtures.

For this mixing, it implies that the scalar glueball  does not
independently exist as a physical state which people explore in
experiment, but the three physical states: $f_0(1370)$, $f_0(1500)$
and $f_0(1710)$ possess glueball components. In fact, many authors
have discussed the mixing of these three physical
states\cite{Anisovich:1996zj,Close:2000yk,Close:2001ga,Giacosa:2005zt,He:2006ij}.
Generally, this issue was discussed based on phenomenology, namely
by fitting data of various reactions, the mixing parameters are
fixed. It would be interesting to investigate this problem from a
more fundamental theory. However, the energy scale for the mixing is
low and the non-perturbative QCD effects may dominate, therefore the
regular perturbative theory does not apply. By contrast, the QCD sum
rules may be the bridge between perturbative quantum field theory
and the non-perturbative phenomena\cite{Shifman:2010zza}, thus
should be a reasonable approach for this research. Two groups have
done the significant work \cite{Narison:1996fm,Narison:1997nw,
Harnett:2008cw}. Narison et al's work fixed the mixing of the three
states: $f_0(1370)$, $f_0(1500)$ and $f_0(1710)$ through the decays
of the light-quark meson and the glueball. In their work, the masses
of the scalar light-quark states and glueball are determined in the
QCD sum rules and by using them to estimate the decay rates of the
corresponding processes they fix the mixing parameters. By contrast,
we assume that the scalar light-quark states $|N,S\rangle$ and the
glueball $|G\rangle$ are un-physical, therefore the masses
independently determined in the QCD sum rules cannot be used to
estimate the decay rates. In another work,  Steele et al. predicted
that the mixing states should involve mixing of $f_0(980)$ with the
$f_0(1500)$ and $f_0(1710)$ in terms of the Gaussian QCD sum rule.
Instead, in our work, we are going to investigate the mixing of the
three states all near $2\mathrm{GeV}$ in the QCD sum rules.

The first step of our work is to define the currents for the
un-physical states: glueball $|G\rangle$, light-quark states
$|N\rangle$ and $|S\rangle$ (N is for u, d quarks, and S is for s
quark), then find their relations to the three physical states:
$|f_1\rangle$, $|f_2\rangle$ and $|f_3\rangle$ via a mixing matrix
$V$.

The work is organized as follows. After this introduction, we calculate
the correlation functions in terms of the QCD sum rules, in Section III, we
formulate the mixing matrix and show the relations between the unphyiscal states and the physical scalar mesons.
In Section IV, we present our numerical results and the last section is devoted to our
conclusion and discussion.

\section{The Correlation Function}

In the scenario of the QCD sum rules, the correlation function $\Pi(q^2)$ is defined as:
\begin{eqnarray}
 \Pi(q^2)=i\int dxe^{iqx}\langle0|T\{J(x),J(0)\}|0\rangle.
\end{eqnarray}
By the dispersion relation, at the hadron hand, the correlation function
can be written as:
\begin{eqnarray}
 \Pi(q^2)={1\over\pi}\int ds{\mathrm{Im}\Pi(s)\over
 s-q^2}.
\end{eqnarray}
After the Borel transformation
and considering the quark-hadron duality, we obtain the ``Moment''
$\mathcal{R}$  as:
\begin{eqnarray}
 \mathcal{R}_k={1\over\pi}\int_0^{s_0}dss^k\mathrm{Im}\Pi(s)e^{-s\tau}
\end{eqnarray}
where $\tau$ is the Boral parameter and $s_0$ is the
threshold for the continuity.

\begin{figure}
\begin{center}
\includegraphics[width=8cm]{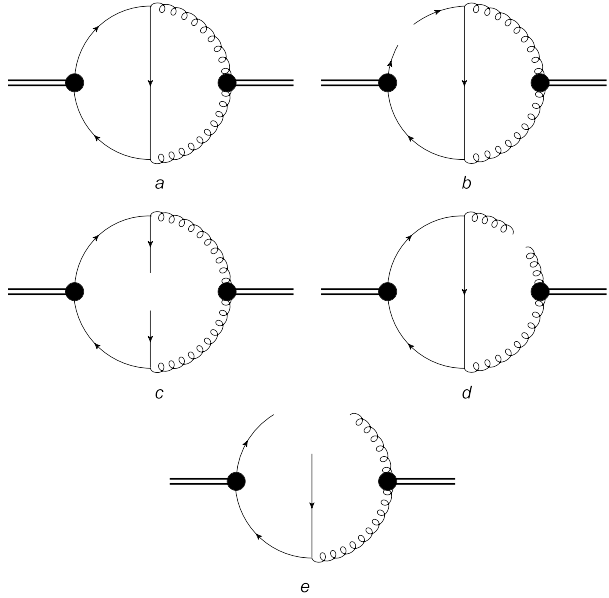}
\caption{ The Feynman Diagrams for $\Pi^\mathrm{QCD}_\mathrm{qg,sg}$
:(a) perturbative part; (b-c) with quark condensates ; (d) with gluon
condensates; (e) with quark-gluon condensates }\label{Mixture-Fey-Loop}
\end{center}
\end{figure}

So in our work, the relevant correlation functions are defined as:
\begin{eqnarray}\label{mixture-cf}
\begin{aligned}
 &\Pi^\mathrm{qq}(q^2)=i\int
dxe^{iqx}\langle0|T\{J_\mathrm{q}(x),J_\mathrm{q}(0)\}|0\rangle\\
 &\Pi^\mathrm{ss}(q^2)=i\int
dxe^{iqx}\langle0|T\{J_\mathrm{s}(x),J_\mathrm{s}(0)\}|0\rangle\\
 &\Pi^\mathrm{gg}(q^2)=i\int
dxe^{iqx}\langle0|T\{J_\mathrm{g}(x),J_\mathrm{g}(0)\}|0\rangle\\
 &\Pi^\mathrm{qg}(q^2)=i\int
dxe^{iqx}\langle0|T\{J_\mathrm{q}(x),J_\mathrm{g}(0)\}|0\rangle\\
 &\Pi^\mathrm{sg}(q^2)=i\int
dxe^{iqx}\langle0|T\{J_\mathrm{s}(x),J_\mathrm{g}(0)\}|0\rangle
\end{aligned}
\end{eqnarray}
\begin{subequations}
where $J_\mathrm{g}(x)$ is
\begin{eqnarray}
 J_\mathrm{g}(x)=\alpha_sG^a_{\mu\nu}(x)G^{a\mu\nu}(x),
\end{eqnarray}
and $J_\mathrm{q,s}(x)$ is:
\begin{eqnarray}
 J_{q,s}(x)=m_{q,s}\psi_{q,s}(x)\bar\psi_{q,s}(x).
\end{eqnarray}
\end{subequations}
The ``Moments'' $\mathcal{R}$ are defined as:
\begin{eqnarray}\label{mixture-R-define}
\begin{aligned}
 &\mathcal{R}_k^\mathrm{qq}={1\over\pi}\int_0^{s_0}dss^k\mathrm{Im}\Pi^\mathrm{qq}(s)e^{-s\tau};\\
 &\mathcal{R}_k^\mathrm{ss}={1\over\pi}\int_0^{s_0}dss^k\mathrm{Im}\Pi^\mathrm{ss}(s)e^{-s\tau};\\
 &\mathcal{R}_k^\mathrm{gg}={1\over\pi}\int_0^{s_0}dss^k\mathrm{Im}\Pi^\mathrm{gg}(s)e^{-s\tau};\\
 &\mathcal{R}_k^\mathrm{qg}={1\over\pi}\int_0^{s_0}dss^k\mathrm{Im}\Pi^\mathrm{qg}(s)e^{-s\tau};\\
 &\mathcal{R}_k^\mathrm{sg}={1\over\pi}\int_0^{s_0}dss^k\mathrm{Im}\Pi^\mathrm{sg}(s)e^{-s\tau},
\end{aligned}
\end{eqnarray}
where $\mathcal{R}_k^\mathrm{gg}$ can be found in
Refs.\cite{Yuan:2009vs,Bagan:1990sy,Huang:1998wj} and
$\mathcal{R}_k^\mathrm{qq,ss}$ is given in
Refs.\cite{Du:2004ki,Reinders:1981ww}. For the mixing current, we
calculate the correlation functions and the ``Moments''
$\mathcal{R}_k^\mathrm{qg,sg}$ are obtained from the Feynman
Diagrams in Fig-\ref{Mixture-Fey-Loop}.

%%%%%%%%%%%%
%%%%%%%%%%%%%

\begin{figure}
\begin{center}
\includegraphics[width=8cm]{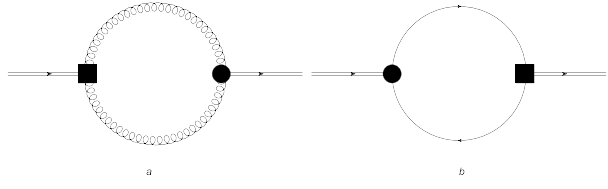}
\caption{ The Counter term for $C_0$, where $\blacksquare$ is for
the vertex correction }\label{Mixture-Fey-Loop-subtract}
\end{center}
\end{figure}

With the Operator Product Expansion (OPE), the correlation function
$\Pi^\mathrm{qq,sg}(q^2)$ is decomposed as:
\begin{eqnarray}\label{mixture-cf-mixing}
 \Pi^\mathrm{qq,sg}(q^2)&=&C_0\hat{O}_0+C_3\langle qq,ss\rangle+C_4\langle \alpha_sG^2\rangle+C_5\langle g_sO_5\rangle+\cdots,
\end{eqnarray}
where $C_i(i=1,3,4,5,\cdots)$ are the Wilson coefficients, and the
operator $\hat{O}_0$ is the unit operator. In the fixed-point
gauge\cite{Pascual:1984zb}, we calculate the two-loop diagram in
Fig-\ref{Mixture-Fey-Loop}(a) and then we have:
\begin{eqnarray}\label{mixture-pert}
 C_0&=&-{1\over\epsilon}{3\alpha_s^2\over2\pi}m_{q,s}^2\log{Q^2\over\nu^2}Q^2+{\alpha_s^2\over\pi^3}m_{q,s}^2Q^2
 \bigg[{3\over2}\log^2{Q^2\over\nu^2}\nonumber\\
 & &+\log{Q^2\over\nu^2}
 \bigg(-3\log4\pi+3\gamma_E-{35\over4}\bigg)\bigg]+\cdots,
\end{eqnarray}
where, $Q^2=-q^2$. In Eq.(\ref{mixture-pert}), we drop out the terms
which are not proportional to $\log[Q^2/\nu^2]$ because they do not
contribute to the moment $\mathcal{R}^\mathrm{q,sg}_k$ and disappear
after the Borel transformation. The Feynman diagrams related to the
counter terms are presented in  Fig-\ref{Mixture-Fey-Loop-subtract}.

%%%%%%%%%%%%
%%%%%%%%%%%%%

In the $\overline{MS}$ scheme, we find:
\begin{subequations}\label{mixture-pert-ct}
\begin{eqnarray}
 C_0^{(a)}=0,
\end{eqnarray}
and
\begin{eqnarray}
 C_0^{(b)}&=&{1\over\epsilon}{3\alpha_s^2\over2\pi^3}m_\mathrm{q,s}^2\log{Q^2\over\nu^2}Q^2
 -{\alpha_s^2\over\pi^3}m_\mathrm{q,s}^2Q^2\bigg[{3\over4}\log^2{Q^2\over\nu^2}\nonumber\\
 & &+\log{Q^2\over\nu^2}\bigg(-3\log4\pi+3\gamma_E-{9\over4}\bigg)\bigg]+\cdots
\end{eqnarray}
\end{subequations}
Eventually
we have the coefficient $C_0$ at the two-loop order as:
\begin{eqnarray}\label{mixture-c0}
 \overline{C}_0&=&C_0+C^{(a)}_0+C^{(b)}_0\nonumber\\
 &=&{\alpha_s^2\over\pi^3}m_\mathrm{q,s}^2Q^2\bigg[{3\over4}\log^2{Q^2\over\nu^2}-{13\over2}\log{Q^2\over\nu^2}\bigg],
\end{eqnarray}
which corresponds to the perturbative contribution to the moments.
The other Wilson coefficients are calculated from
Fig-\ref{Mixture-Fey-Loop}(b-e) as:
\begin{eqnarray}
\begin{aligned}
 &C_3=-4\pi{\alpha^2\over\pi^2}m_\mathrm{q,s}\log{Q^2\over\nu^2};\\
 &C_4={m_\mathrm{q,s}^2\over
 Q^2}\bigg[-{\alpha_s\over\pi}\log{Q^2\over\nu^2}+3{\alpha_s\over\pi}\bigg];\\
 &C_5=-{2\over Q^2}m_\mathrm{q,s}\alpha_s.
\end{aligned}
\end{eqnarray}
With the correlation function Eq.(\ref{mixture-cf-mixing}), the
moment is:
\begin{subequations}\label{mixture-R-qg-qcd}
\begin{eqnarray}
 \mathcal{R}^\mathrm{qg,sg}_0&=&{1\over\tau^2}(1-\rho_1(s_0\tau))a_0^\mathrm{q,s}
 -{2a_1^\mathrm{q,s}\over\tau^2}\bigg[\gamma_E+E_1(s_0\tau)+\log
 s_0\tau+e^{-s_0}\tau-1\nonumber\\
 &
 &-(1-\rho_1(s_0\tau))\log{s_0\over\nu^2}\bigg]-{b_1^\mathrm{q,s}\over\tau}(1-\rho_0(s_0\tau))m_q\langle
 q\bar{q}\rangle\nonumber\\
 & &+\bigg[c_0^\mathrm{q,s}-c_1^\mathrm{q,s}\big(\gamma_E+\log\tau\nu^2+E_1(s_0\tau)\big)\bigg]\langle\alpha_sG^2\rangle+d_0^\mathrm{q,s}\langle
 g_sO_5\rangle\nonumber\\
\end{eqnarray}
where,
\begin{eqnarray}
\begin{aligned}
 &a_0^\mathrm{q,s}=-{13\alpha_s^2\over4\pi^3}m_{q,s}^2 & a_1^\mathrm{q,s}&={3\alpha_s^2\over4\pi^3}m_{q,s}^2\\
 &b_1^\mathrm{q,s}=-4{\alpha_s^2\over\pi} & c_0^\mathrm{q,s}&={3\alpha_s\over\pi}m_{q,s}\\
 &c_1^\mathrm{q,s}=-{\alpha_s\over\pi}m_{q,s}^2 & d_0^\mathrm{q,s}&=-2\alpha_sm_{q,s}
\end{aligned}
\end{eqnarray}
and $\rho_{1,2\cdots}(x)$ and $E_1(x)$ are already given in
\cite{Yuan:2009vs,Bagan:1990sy,Huang:1998wj}.
\end{subequations}

It is noted that our result is different from that given in
\cite{Harnett:2008cw}. This is understood since different
subtraction schemes are employed in the two works.

\section{Equations for Mixing Matrix $V$}
We define the physical states as $|f_1\rangle$£¬$|f_2\rangle$ and
$|f_2\rangle$, whereas the un-physical states as
$|N\rangle=|\bar{q}q\rangle$, $|S\rangle=|\bar{s}s\rangle$ and
$|G\rangle$. The mixing matrix connecting them is:
\begin{eqnarray}\label{mixture-state}
\begin{pmatrix}
 |f_1\rangle\\
 |f_2\rangle\\
 |f_3\rangle\\
\end{pmatrix}
=
\begin{pmatrix}
 V_{11} & V_{12} & V_{13}\\
 V_{21} & V_{22} & V_{23}\\
 V_{31} & V_{32} & V_{33}
\end{pmatrix}
\begin{pmatrix}
 |N\rangle\\
 |S\rangle\\
 |G\rangle\\
\end{pmatrix}
\end{eqnarray}
According to the first approximation, it is assumed that
$|N\rangle$, $|S\rangle$ and $|G\rangle$ constitute a complete
basis\cite{Close:2000yk}, but as a matter of fact, when the other
resonances $f_0(1790)$ and $f_0(1812)$ were observed by the BES
collaboration \cite{Ablikim:2004wn,Ablikim:2006dw}, we suggested
that the hybrids might join the game and mix with the aforementioned
states \cite{He:2006ij,He:2006tw}. But it seems that one can first
ignore the hybrids which might be heavier than the other three, and
assume that the three physical mesons are only composed of the regular quark
and glueball components. We will discuss this issue in the last
section. So, the mixing matrix $V$ transforms the flavor
representation into the physical representation, i.e. the mass
representation, so it must be unitary, thus we have:
\begin{eqnarray}\label{mixture-normalization}
\begin{array}{l}
 V_{11}^2+V_{21}^2+V_{31}^2=1;\\
 V_{12}^2+V_{22}^2+V_{32}^2=1;\\
 V_{13}^2+V_{23}^2+V_{33}^2=1,
\end{array}
\end{eqnarray}
and the conditions are enforced
\begin{eqnarray}\label{mixture-orthogonal}
\begin{array}{l}
 V_{11}V_{12}+V_{21}V_{22}+V_{31}V_{32}=0;\\
 V_{11}V_{13}+V_{21}V_{23}+V_{31}V_{33}=0;\\
 V_{11}V_{13}+V_{21}V_{23}+V_{31}V_{33}=0.
 \end{array}
\end{eqnarray}
Next, we will build the equations to solve this mixing matrix $V$ in
terms of the QCD sum rules.

In QCD sum rules, the integrand of the dispersion integral includes
the imaginary part of the correlation function $\Pi$ at $q^2>0$ and
then one inserts a complete set of physical states of $0^{++}$
hadrons between the currents\cite{Colangelo:2000dp}. In the
quark-hadron duality the lowest states' contributions dominate and
the contributions of the higher exited states and the continuum
should be dropped out by introducing the threshold $s_0$ as the
lower bound of the integration. Since we are investigating the
mixing, we insert all the three lowest states $|f_1\rangle$,
$|f_2\rangle$ and $|f_3\rangle$ into Eq.(\ref{mixture-R-define}) and
then we have:
\begin{eqnarray}\label{mixture-Im-f123}
 {1\over\pi}\mathrm{Im}\Pi^\mathrm{ij}(s)&=&\sum_{n=1,2,3}\langle0|J_\mathrm{i}|f_n\rangle\langle
 f_n|J_\mathrm{j}|0\rangle\delta(s-m_n^2)+\rho^h(s)\theta(s-s_0^h),
\end{eqnarray}
where $\mathrm{i,j}=\mathrm{q,s,g}$ stand for the different currents
(see Eq.(\ref{mixture-cf})), $n$ labels the state in the complete
set, $\rho^h(s)$ represents all the higher exited states and the
continuum and $s_0^h$ is the threshold for these higher states.

Putting Eq.(\ref{mixture-Im-f123}) back into
Eq.(\ref{mixture-R-define}) and with the quark-hadron duality, we
finally have the moments as:
\begin{subequations}\label{mixture-R-hadron}
\begin{eqnarray}
 \mathcal{R}^\mathrm{qq}_0&=&\langle0|J_\mathrm{q}|f_1\rangle^2e^{-m_1^2\tau}
 +\langle0|J_\mathrm{q}|f_2\rangle^2e^{-m_2^2\tau}
 +\langle0|J_\mathrm{q}|f_3\rangle^2e^{-m_3^2\tau}\nonumber\\
 &=&\bigg(V_{11}^2e^{-m_1^2\tau}+V_{21}^2e^{-m_2^2\tau}+V_{31}^2e^{-m_3^2\tau}\bigg)
 \langle0|J_\mathrm{q}|N\rangle^2
\end{eqnarray}
\begin{eqnarray}
 \mathcal{R}^\mathrm{ss}_0&=&\langle0|J_\mathrm{s}|f_1\rangle^2e^{-m_1^2\tau}
 +\langle0|J_\mathrm{s}|f_2\rangle^2e^{-m_2^2\tau}
 +\langle0|J_\mathrm{s}|f_3\rangle^2e^{-m_3^2\tau}\nonumber\\
 &=&\bigg(V_{12}^2e^{-m_1^2\tau}+V_{22}^2e^{-m_2^2\tau}+V_{32}^2e^{-m_3^2\tau}\bigg)
 \langle0|J_\mathrm{s}|S\rangle^2
\end{eqnarray}
\begin{eqnarray}
 \mathcal{R}^\mathrm{gg}_0&=&\langle0|J_\mathrm{g}|f_1\rangle^2e^{-m_1^2\tau}
 +\langle0|J_\mathrm{g}|f_2\rangle^2e^{-m_2^2\tau}
 +\langle0|J_\mathrm{g}|f_3\rangle^2e^{-m_3^2\tau}\nonumber\\
 &=&\bigg(V_{13}^2e^{-m_1^2\tau}+V_{23}^2e^{-m_2^2\tau}+V_{33}^2e^{-m_3^2\tau}\bigg)
 \langle0|J_\mathrm{g}|G\rangle^2
\end{eqnarray}
\begin{eqnarray}
 \mathcal{R}^\mathrm{qg}_0&=&\langle0|J_\mathrm{q}|f_1\rangle\langle f_1|J_\mathrm{g}|0\rangle e^{-m_1^2\tau}
 +\langle0|J_\mathrm{q}|f_2\rangle\langle f_2|J_\mathrm{g}|0\rangle
 e^{-m_2^2\tau}\nonumber\\
 & &+\langle0|J_\mathrm{q}|f_3\rangle\langle f_3|J_\mathrm{g}|0\rangle e^{-m_3^2\tau} \nonumber\\
 &=&\bigg(V_{11}V_{13}e^{-m_1^2\tau}+V_{21}V_{23}e^{-m_2^2\tau}+V_{31}V_{33}e^{-m_3^2\tau}\bigg)\langle0|J_\mathrm{q}|N\rangle\langle
 G|J_\mathrm{g}|0\rangle
\end{eqnarray}
\begin{eqnarray}
 \mathcal{R}^\mathrm{sg}_0&=&\langle0|J_\mathrm{s}|f_1\rangle\langle f_1|J_\mathrm{g}|0\rangle e^{-m_1^2\tau}
 +\langle0|J_\mathrm{s}|f_2\rangle\langle f_2|J_\mathrm{g}|0\rangle
 e^{-m_2^2\tau}\nonumber\\
 & &+\langle0|J_\mathrm{s}|f_3\rangle\langle f_3|J_\mathrm{g}|0\rangle e^{-m_3^2\tau} \nonumber\\
 &=&\bigg(V_{12}V_{13}e^{-m_1^2\tau}+V_{22}V_{23}e^{-m_2^2\tau}+V_{32}V_{33}e^{-m_3^2\tau}\bigg)\langle0|J_\mathrm{s}|S\rangle\langle
 G|J_\mathrm{g}|0\rangle,
\end{eqnarray}
\end{subequations}
where $m_1$, $m_2$ and $m_3$ are the masses of  $|f_1\rangle$,
$|f_2\rangle$ and $|f_3\rangle$. In Eq.(\ref{mixture-R-hadron}), and
the relationship of the physical and the un-physical states is
involved in the calculations: such the concerned current only
couples to the certain un-physical state with the right quantum
number and flavor. For example, the current of the glueball cannot
couple to the state of the light-quark, vice versa. For the
physical state $|f_1\rangle$, $|f_2\rangle$ and $|f_3\rangle$, we
have:
\begin{eqnarray}
\left\{
\begin{aligned}
 &\langle0|J_q|f_i\rangle=\langle0|J_q|V_{i1}|N\rangle=V_{i1}\langle0|J_q|N\rangle~;\\
 &\langle0|J_s|f_i\rangle=\langle0|J_s|V_{i2}|S\rangle=V_{i2}\langle0|J_s|S\rangle~;\\
 &\langle0|J_g|f_i\rangle=\langle0|J_g|V_{i3}|G\rangle=V_{i3}\langle0|J_g|G\rangle~,\\
\end{aligned}
\right.
\end{eqnarray}
where $i=1,2,3$ for the three physical states. The un-physical
states $|N\rangle$, $|S\rangle$ and $|G\rangle$ directly couple to
the certain currents, but do not correspond to any physical values.
Thus we need to relate them to the physical states in terms via the
moments in Eq. (\ref{mixture-R-hadron}). Thus we are able to
establish the equations for the ratios among the moments:
\begin{widetext}
\begin{eqnarray}\label{mixture-eqn}
\begin{aligned}
 {\mathcal{R}^\mathrm{qq}_{k+1}(\tau,s^\mathrm{qq}_0)\over\mathcal{R}^\mathrm{qq}_k(\tau,s^\mathrm{qq}_0)}
 &={V_{11}^2e^{-m_1^2\tau}m_1^{2(k+1)}+V_{21}^2e^{-m_2^2\tau}m_2^{2(k+1)}+V_{31}^2e^{-m_3^2\tau}m_3^{2(k+1)}\over
 V_{11}^2e^{-m_1^2\tau}m_1^{2k}+V_{21}^2e^{-m_2^2\tau}m_2^{2k}+V_{31}^2e^{-m_3^2\tau}m_3^{2k}}~;\\
 {\mathcal{R}^\mathrm{ss}_{k+1}(\tau,s^\mathrm{ss}_0)\over\mathcal{R}^\mathrm{ss}_k(\tau,s^\mathrm{ss}_0)}
 &={V_{12}^2e^{-m_1^2\tau}m_1^{2(k+1)}+V_{22}^2e^{-m_2^2\tau}m_2^{2(k+1)}+V_{32}^2e^{-m_3^2\tau}m_3^{2(k+1)}\over
 V_{12}^2e^{-m_1^2\tau}m_1^{2k}+V_{22}^2e^{-m_2^2\tau}m_2^{2k}+V_{32}^2e^{-m_3^2\tau}m_3^{2k}}~;\\
 {\mathcal{R}^\mathrm{gg}_{k+1}(\tau,s^\mathrm{gg}_0)\over\mathcal{R}^\mathrm{gg}_k(\tau,s^\mathrm{gg}_0)}
 &={V_{13}^2e^{-m_1^2\tau}m_1^{2(k+1)}+V_{23}^2e^{-m_2^2\tau}m_2^{2(k+1)}+V_{33}^2e^{-m_3^2\tau}m_3^{2(k+1)}\over
 V_{13}^2e^{-m_1^2\tau}m_1^{2k}+V_{23}^2e^{-m_2^2\tau}m_2^{2k}+V_{33}^2e^{-m_3^2\tau}m_3^{2k}}~;\\
 {\mathcal{R}^\mathrm{qg}_{k+1}(\tau,s^\mathrm{qg}_0)\over\mathcal{R}^\mathrm{qg}_k(\tau,s^\mathrm{qg}_0)}
 &={V_{11}V_{13}e^{-m_1^2\tau}m_1^{2(k+1)}+V_{21}V_{23}e^{-m_2^2\tau}m_2^{2(k+1)}+V_{31}V_{33}e^{-m_3^2\tau}m_3^{2(k+1)}\over
 V_{11}V_{13}e^{-m_1^2\tau}m_1^{2k}+V_{21}V_{23}e^{-m_2^2\tau}m_2^{2k}+V_{31}V_{33}e^{-m_3^2\tau}m_3^{2k}}~;\\
 {\mathcal{R}^\mathrm{sg}_{k+1}(\tau,s^\mathrm{sg}_0)\over\mathcal{R}^\mathrm{sg}_k(\tau,s^\mathrm{sg}_0)}
 &={V_{12}V_{13}e^{-m_1^2\tau}m_1^{2(k+1)}+V_{22}V_{23}e^{-m_2^2\tau}m_2^{2(k+1)}+V_{32}V_{33}e^{-m_3^2\tau}m_3^{2(k+1)}\over
 V_{12}V_{13}e^{-m_1^2\tau}m_1^{2k}+V_{22}V_{23}e^{-m_2^2\tau}m_2^{2k}+V_{32}V_{33}e^{-m_3^2\tau}m_3^{2k}}~.
\end{aligned}
\end{eqnarray}
\end{widetext}
Totally we have eight equations in Eq.(\ref{mixture-eqn}) and
Eq.(\ref{mixture-normalization}) for determining the mixing matrix.
Supposing the matrix is real, there should be nine independent
elements, but we only have eight equations, so that this equation
group is not enough to directly determine the whole matrix. However, as
we know, the matrix is unitary (as the matrix is real as assumed, it
is an orthogonal matrix), thus we may gain an extra equation to fix
all elements of the matrix. Namely, on the other hand, if we fix one
element of the matrix $V$, in our work, for example,  $V_{23}$,
then all other elements of the matrix $V$ can be obtained by solving
these eight equations. Sequently, let $V_{23}$ run in the region
$[-1,1]$, the unitarity condition may help to eventually fix its
value and the best fitting of the $V$ is expected.

\section{numerical results}

In Eq.(\ref{mixture-eqn}), it needs the values of the condensates
and some other parameters  as inputs. From \cite{Colangelo:2000dp},
we set them as:
\begin{eqnarray}
\begin{aligned}
 &m_q=0.008\mathrm{GeV}, & &m_s=0.14\mathrm{GeV},\\
 &m_0=\sqrt{0.8}\mathrm{GeV}, & &\langle\bar
 qq\rangle=-0.24^3\mathrm{GeV}^3,\\
 &\langle\alpha_sG^2\rangle=0.06\mathrm{GeV}^4, & &\langle
 g_sO_5\rangle=m_0^2\langle\bar qq\rangle\mathrm{GeV}^5.
\end{aligned}
\end{eqnarray}

%%%%%%%%%%%%%%%%%%%%%
\begin{figure}
\begin{center}
\includegraphics[width=15cm]{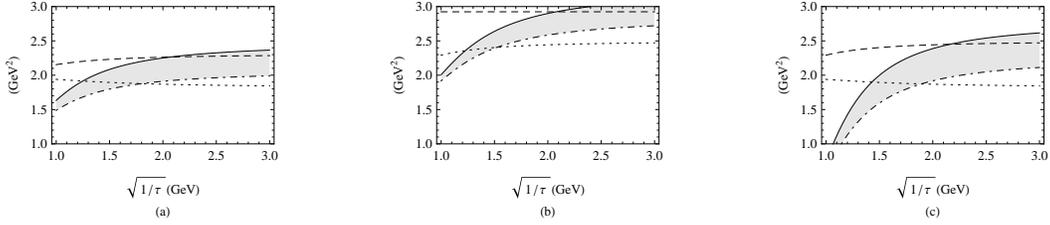}
\caption{ The Gray Area is the value of the lhs. of the first three
equations in Eq.(\ref{mixture-eqn}):(a)
$s_{0\mathrm{qq}}\in[3.1,3.7]\mathrm{GeV}^2$; (b)
$s_{0\mathrm{ss}}\in[4.2,4.8]\mathrm{GeV}^2$;(c)
$s_{0\mathrm{gg}}\in[3.5,4.1]\mathrm{GeV}^2$. The Area between the
Dashed line and the Dotted line is the value of the rhs. of the
first three equations in Eq.(\ref{mixture-eqn}): (a)
$|V_{11,21}|\in[0.6,0.8]$; (b) $|V_{12,22}|\in[0.01,0.5]$; (c)
$|V_{13,23}|\in[0.5,0.8]$. So the overlapping region is the proper
parameter area for the mixing matrix $V$. }\label{Mixture-para-1}
\end{center}
\end{figure}
\begin{figure}
\begin{center}
\includegraphics[width=10cm]{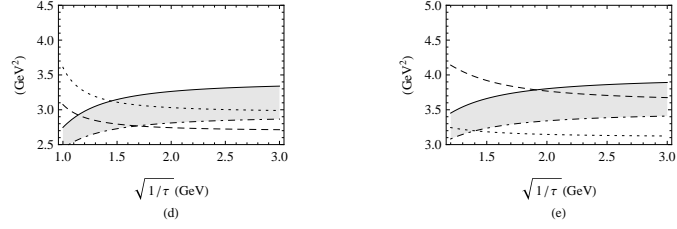}
\caption{ The Gray Area is the value of the lhs. of the last two
equations in Eq.(\ref{mixture-eqn}):(d)
$s_{0\mathrm{qg}}\in[3.6,4.2]\mathrm{GeV}^2$; (e)
$s_{0\mathrm{sg}}\in[4.2,4.8]\mathrm{GeV}^2$. The Area between the
Dashed line and the Dotted line is the value of the rhs. of the last
two equations in Eq.(\ref{mixture-eqn}): (d)
$V_{11,21}\in[-0.7,-0.6]$, $V_{31}\in[0.6,0.72]$, and
$V_{32}\in[-0.72,-0.6]$; (e) $V_{21}\in[0.1,0.3]$,
$V_{22}\in[0.45,0.47]$, $V_{31}\in[0.6,0.7]$ and
$V_{32}\in[-0.69,-0.60]$}\label{Mixture-para-2}
\end{center}
\end{figure}

%%%%%%%%%%%%%%%%%%%%
The other parameters are related to the QCD sum rules: the Borel
parameter $\tau$ and the threshold of $s_0^\mathrm{qq}$,
$s_0^\mathrm{ss}$, $s_0^\mathrm{gg}$, $s_0^\mathrm{qg}$ and
$s_0^\mathrm{sq}$ defined in Eq.(\ref{mixture-eqn}). By the general
strategy, one should search for plateaus in the diagrams of the
correlation versus the Borel parameter and the threshold $s_0$. Only
the parameters fall in a certain region, the plateaus can appear,
namely within the plateaus the results are not sensitive to the
choice of Borel parameter and $s_0$, then are trustworthy. In this
work, there are six correlation functions in total, so we require
all of them to have a common plateau region for the Borel parameter,
where all the six moments are relatively independent of the Borel
parameter. Obviously this condition is not easy to be satisfied.
Once such a region is found, we would be able to conclude that the
results based on the QCD sum rules make sense. The dependence of all
six moments on the Borel parameter are presented in
Fig-\ref{Mixture-para-1} and Fig-\ref{Mixture-para-2}. And we can
see obvious appearance of plateaus.

%%%%%%%%%%%%%%%%%%%%%
%%%%%%%%%%%%%%%%%%%%%

We first have to check if in the parameter regions
Eqs.(\ref{mixture-eqn}) have real solutions. We find that there are
indeed. As we require the matrix $V$ to be real,
only a very narrow parameter space is available. The
Fig-\ref{Mixture-para-1} and Fig-\ref{Mixture-para-2} show the
values of the right-hand side (rhs) and the left-hand side (lhs) of
the equations in Eq.(\ref{mixture-eqn}), where the
Fig-\ref{Mixture-para-1} is for the first three equations, and the
Fig-\ref{Mixture-para-2} is for the last two equations. Taking the error tolerance into account,
the lines would be widened into bands, in the
Fig-\ref{Mixture-para-1} and Fig-\ref{Mixture-para-2}, the region
between the Dashed line and the Dotted line is for the rhs of the
Eq.(\ref{mixture-eqn}) and the Gray one is for lhs. It is clear
that, only in the overlapping region, rhs and lhs can be equal, and
appearance of the overlapping region implies that a solution of the
Eq.(\ref{mixture-eqn}) may exist.

Searching for such an overlapping region in Fig-\ref{Mixture-para-1} and
Fig-\ref{Mixture-para-2}, one needs to find a proper
parameter space. Eventually, we have found a satisfactory region where the
best-fitted parameters are: the Borel parameter
$\tau\in[1/1.8^2,1/2.1^2]\mathrm{GeV}^{-2}$ and the five thresholds which
must be close to  $s_0^\mathrm{qq}=3.4\mathrm{GeV}^2$, are
$s_0^\mathrm{ss}=4.5\mathrm{GeV}^2$,
$s_0^\mathrm{gg}=3.8\mathrm{GeV}^2$,
$s_0^\mathrm{qg}=3.9\mathrm{GeV}^2$ and
$s_0^\mathrm{sq}=4.5\mathrm{GeV}^2$.
At the same time, the allowed
value ranges of the matrix elements $V_{ij}$ are also set. From
Fig-\ref{Mixture-para-1}, one notices that only as the matrix
elements fall in the following regions:
\begin{eqnarray}\label{Mixture-para}
\begin{aligned}
 &V_{11}\in[-0.6,-0.8] & &V_{12}\in[0.01,0.5] & &V_{13}\in[0.5,0.8]\\
 &V_{21}\in[-0.6,-0.8] & &V_{22}\in[0.3,0.6] & &V_{23}\in[0.5,0.8]\,
\end{aligned}\nonumber\\
\end{eqnarray}
all the requirements are satisfied.
It is also noted that due to the unitarity condition
(\ref{mixture-normalization}), $V_{3i}$ ( $i=1,~2,~3$)  depend on
other elements $V_{1i}$ and $V_{2i}$, thus their value-ranges would
be uniquely determined (there might be a sign difference), once the
others are fixed.

%So here we know that, only in the narrow parameter space for
%parameter and the region of the matrix element (\ref{Mixture-para}),
%the real solution of the Eq.(\ref{mixture-eqn}) can exist, since
%they have the overlapping region in Fig-\ref{Mixture-para-1} and
%Fig-\ref{Mixture-para-2}. If the parameters do not fall in this
%region, the solution of the Eq.(\ref{mixture-eqn}) cannot be real or
%the orthogonality in Eq.(\ref{mixture-orthogonal}) is not satisfied.

Fig-\ref{Mixture-para-2} corresponds to the last two equations
in Eq.(\ref{mixture-eqn}), and apparently overlapping regions
exist when the matrix elements of $V$ reside in the ranges (\ref{Mixture-para}).
Moreover, for the last two equations in Eq.(\ref{mixture-eqn}), we set $k=3$.
The reason is that, only when $k=0~\mathrm{or}~3$, the equations
Eq.(\ref{mixture-eqn})  have real solutions. However, $k=0$ is not
proper since when $k=0$, the lhs does not appear in the plateau.

%By the way, one thing must be noted here. It is possible to have the
%overlapping region with other value of the parameters actually. But
%as what we do next, we solve the Eq.(\ref{mixture-eqn}) numerically,
%and we find that, considering the real and unitary matrix $V$ we
%need to solve, the other parameters can not satisfy the requirement,
%and then we drop them all.

\begin{figure}
\begin{center}
\includegraphics[width=15cm]{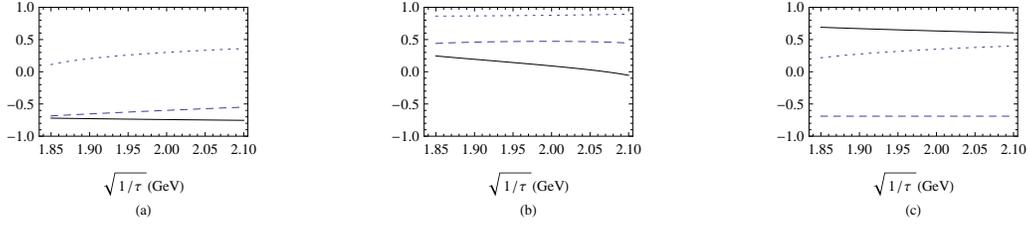}
\caption{ The dependence of the matrix elements $V_{ij}$ on the
Borel parameter $\tau$: in (a,b,c) the line is for $V_{11,12,13}$,
the dashed line is for $V_{21,22,23}$ and the Dotted line is for
$V_{31,32,33}$ }\label{mixture-fig-vmatrix&boral}
\end{center}
\end{figure}

We solve the equations Eq.(\ref{mixture-eqn}) together with the
three equations in Eq.(\ref{mixture-normalization}). Our strategy is
to set $V_{23}$ as a free parameter and let it run within a range.
We find that only when $V_{23}\sim -0.69$, the matrix $V$ is real
and orthogonal. The numerical solution is given in
Tab-\ref{mixture-vmatrix&boral} and the dependence of the matrix
elements $V_{ij}$ on the Borel parameter $\tau$ is shown in
Fig-\ref{mixture-fig-vmatrix&boral}.

Since such terms $V^2_{ij}$ exist in Eq.(\ref{mixture-eqn}), the
solution may not be unique. As a matter of fact, we obtain eight
independent groups of solutions. However, enforcing the unitary condition
to the matrix $V$, we find that several groups are
practically identical (i.e. they deviate from each other by just a common phase)
and others must be dropped out because they do not satisfy the orthogonal condition. Finally
only one group of solutions remains which is presented in the
following table.

%%%%%%%%%%%%%%%%%%%%%%%
%%%%%%%%%%%%%%%%%%%%%%%

%%%%%%%%%%%%%%%%%%%%%%%%%%%%%%%%%%%%%%%%%%%%%%%%%%%%%%%%%%%%%%%%%%%
\begin{center}
\begin{table}[!h]
\caption{The relationship of the elements of the $V$ and $\tau$
}\label{mixture-vmatrix&boral} \vspace{0.5cm}
\begin{tabular}{ccccccccc}
\hline\hline
 $\sqrt{1/\tau}$(GeV) & 1.8 & 1.85 & 1.9 & 1.95 & 2.0 & 2.05 & 2.1 & 2.15
\\
\hline
 $V_{11}$ & -0.71 & -0.72 & -0.73 & -0.74 & -0.74 & -0.75 & -0.75 & -0.76
\\
 $V_{21}$ & -0.72 & -0.69 & -0.65 & -0.63 & -0.60 & -0.57 & -0.55 & -0.52
\\
 $V_{31}$ & 0.01 & 0.11 & 0.20 & 0.26 & 0.30 & 0.33 & 0.36 & 0.38
\\
 $V_{12}$ & 0.31 & 0.25 & 0.19 & 0.14 & 0.09 & 0.03 & -0.06 & -0.15
\\
 $V_{22}$ & 0.40 & 0.44 & 0.46 & 0.47 & 0.47 & 0.47 & 0.44 & 0.41
\\
 $V_{32}$ & 0.87 & 0.86 & 0.87 & 0.87 & 0.88 & 0.88 & 0.89 & 0.89
\\
 $V_{13}$ & 0.71 & 0.69 & 0.67 & 0.65 & 0.63 & 0.62 & 0.60 & 0.59
\\
 $V_{23}$ & -0.69 & -0.69 & -0.69 & -0.69 & -0.69 & -0.69 & -0.69 & -0.69
\\
 $V_{33}$ & 0.12 & 0.22 & 0.27 & 0.32 & 0.35 & 0.38 & 0.40 & 0.42
\\ \hline\hline
\end{tabular}
\end{table}
\end{center}
%%%%%%%%%%%%%%%%%%%%%%%%%%%%%%%%%%%%%%%%%%%%%%%%%%%%%%%%%%%%%%%%%%%

Our numerical results show that $\tau=1/2^2\mathrm{GeV}^2$ is the
center of the common plateau, and the mixing matrix $V$ is
\begin{eqnarray}\label{mixture-vmatrix}
V=
\begin{pmatrix}
 -0.74^{+0.02}_{-0.02} & 0.09^{+0.22}_{-0.26} &
 0.63^{+0.08}_{-0.04}\\
 -0.60^{+0.08}_{-0.08} & 0.47^{+0.03}_{-0.05} & -0.69^*\\
 0.30^{+0.08}_{-0.19} & 0.88^{+0.02}_{-0.02} & 0.35^{+0.07}_{-0.13}
\end{pmatrix}
\end{eqnarray}

The numerical analysis indicates that the matrix elements $V_{11}$,
$V_{21}$, $V_{22}$, $V_{32}$ and $V_{13}$ do not change much when
the Borel parameter runs from $1/1.8^2\mathrm{GeV}^{-2}$ to
$1/2.15^2\mathrm{GeV}^{-2}$, but it is also noted that  the errors
of $V_{31}$, $V_{12}$ and $V_{33}$ are relatively larger.

The ratio of the contribution  of the perturbative part to the
``Moments'' $\mathcal{R}$ and the lowest state below the threshold
is given in Fig-\ref{Mixture-ratio}.

\section{Conclusion and discussion}

%%%%%%%%%%%%
\begin{figure*}
\begin{center}
\includegraphics[width=10cm]{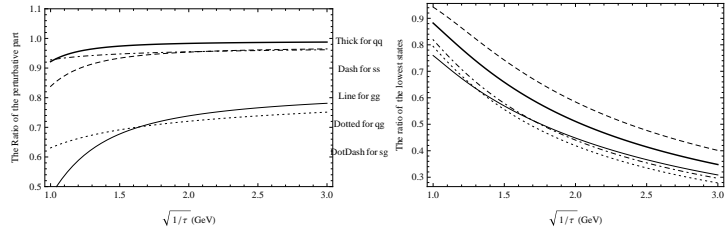}
\caption{ The ratio of the contribution  of the perturbative part to
the ``Moments'' $\mathcal{R}$ and the lowest state under the
threshold point } \label{Mixture-ratio}
\end{center}
\end{figure*}
%%%%%%%%%%%%%

From the FIG-\ref{Mixture-ratio} we find that, within the range of
$\tau\in[1/2.15^2,1/1.8^2]\mathrm{GeV}^{-2}$, the fraction of the
perturbative part in the total contribution is over $60\%$. By the
general principle of the QCD sum rules, after performing the Borel
transformation, the perturbative contribution should dominate, and
it is a criterion for judging the reliability of the results. 60\%
is not too bad at all.
%On the other hand, the ratio of the lowest
%states under the threshold point is about 50\%. Given the constrain
%of the matrix $V$, 50\% is acceptable.

For a comparison let us write down the mixing matrix given by Close
et al. \cite{Close:2000yk}:
\begin{eqnarray}
V_{C}=
\begin{pmatrix}
 -0.79 & -0.13 & 0.60\\
 -0.62 & 0.37 & -0.69\\
 0.14 & 0.91 & 0.39
\end{pmatrix}
\end{eqnarray}

In this work, we calculate the mixing of the $|N\rangle$,
$|S\rangle$ and $|G\rangle$ to result in the physical resonances
$f_0(1370)$, $f_0(1500)$ and $f_0(1710)$. The mixing matrix
Eq.(\ref{mixture-vmatrix}) which we obtained in the QCD sum rules is
consistent with $V_C$ \cite{Close:2000yk} which was achieved based
on phenomenological studies.

This work is based on the conjecture of Close and
Kirk\cite{Close:2000yk} that only the mesons heavier than
$1\mathrm{GeV}$ are mixtures of $q\bar q$, $s\bar s$ and glueball
$G$, because the lattice results indicate that the mass of $0^{++}$
is around $1.5\sim1.7\mathrm{GeV}$. This was also suggested by
Narison et al. in their earlier
papers\cite{Narison:1984hu,Narison:1988ts,Mathieu:2008me,Shuiguo:2010ak}.

With this picture we calculate the mixing off-diagonal correlators
which result in the physical resonances $f_0(1370)$, $f_0(1500)$ and
$f_0(1710)$. Narison and his collaborators computed the off-diagonal
correlators  in their pioneer work when they considered a mixing
between meson and glueball\cite{Narison:1984bv,Mennessier:1985mz}.

Today, thanks to the progress of experimental facilities and
innovation of the data-analysis, many new resonances have been
observed and data are updated. The available new data enable us to
re-study the mixing effects, even though the basic techniques have
been provided in those pioneer papers. That is the aim of this work.
We are indeed very encouraged by the consistency between the
numerical results obtained in terms of the QCD sum rules and that
gained by the phenomenological research. It implies that the QCD sum
rules are really a good approach for studying hadron physics even
though certain uncertainties unavoidably exist.

Moreover, as we indicated above, the another two resonances
$f_0(1790)$ and $f_0(1812)$ were observed and they also reside in
the range of 1 to 2 GeV, therefore we do not have reason to ignore a
possibility that all the five physical states $f_0(1370)$,
$f_0(1500)$, $f_0(1710)$, $f_0(1790)$ and $f_0(1812)$ are mixtures
of $q\bar q$, $s\bar s$, $q\bar q G$ and $s\bar s G$ and glueball of
$0^{++}$. But it would be much more difficult to calculate the
mixing not only because then we have to deal with a five-dimensional
matrix, but also the leading order of the perturbative part of the
correlation function is two-loop feynman diagrams. But if it is the
real physics, we need to carry out the calculations, and it will be
the task of our next work.

\vspace{1cm}

\section*{Acknowledgments}

This project is supported by the National Natural Science Foundation
of China (NSFC) under Contracts No. 10775073 and No. 11005079.


\vspace{1cm}

\begin{thebibliography}{}

%\cite{Ishikawa:1982tb}
\bibitem{Ishikawa:1982tb}
  K.~Ishikawa, M.~Teper and G.~Schierholz,
  %``THE GLUEBALL MASS SPECTRUM IN QCD: FIRST RESULTS OF A LATTICE MONTE CARLO
  %CALCULATION,''
  Phys.\ Lett.\  B {\bf 110}, 399 (1982).
  %%CITATION = PHLTA,B110,399;%%

%\cite{Berg:1982kp}
\bibitem{Berg:1982kp}
  B.~Berg and A.~Billoire,
  %``Glueball Spectroscopy in Four-Dimensional SU(3) Lattice Gauge Theory. 1,''
  Nucl.\ Phys.\  B {\bf 221}, 109 (1983).
  %%CITATION = NUPHA,B221,109;%%

%\cite{de Forcrand:1984rs}
\bibitem{de Forcrand:1984rs}
  P.~de Forcrand, G.~Schierholz, H.~Schneider and M.~Teper,
  %``THE 0++ GLUEBALL MASS IN SU(3) LATTICE GAUGE THEORY: TOWARDS DEFINITIVE
  %RESULTS,''
  Phys.\ Lett.\  B {\bf 152}, 107 (1985).
  %%CITATION = PHLTA,B152,107;%%

%\cite{Teper:1987wt}
\bibitem{Teper:1987wt}
  M.~Teper,
  %``An Improved Method for Lattice Glueball Calculations,''
  Phys.\ Lett.\  B {\bf 183}, 345 (1987).
  %%CITATION = PHLTA,B183,345;%%

%\cite{Albanese:1987ds}
\bibitem{Albanese:1987ds}
  M.~Albanese {\it et al.}  [APE Collaboration],
  %``Glueball Masses and String Tension in Lattice QCD,''
  Phys.\ Lett.\  B {\bf 192}, 163 (1987).
  %%CITATION = PHLTA,B192,163;%%

%\cite{Teper:1998kw}
\bibitem{Teper:1998kw}
  M.~J.~Teper,
  %``Glueball masses and other physical properties of SU(N) gauge theories in D
  %= (3+1): A Review of lattice results for theorists,''
  arXiv:hep-th/9812187.
  %%CITATION = HEP-TH/9812187;%%

%\cite{Morningstar:1999rf}
\bibitem{Morningstar:1999rf}
  C.~J.~Morningstar and M.~J.~Peardon,
  %``The Glueball spectrum from an anisotropic lattice study,''
  Phys.\ Rev.\  D {\bf 60}, 034509 (1999)
  [arXiv:hep-lat/9901004].
  %%CITATION = PHRVA,D60,034509;%%

%\cite{Ishii:2002ww}
\bibitem{Ishii:2002ww}
  N.~Ishii, H.~Suganuma and H.~Matsufuru,
  %``Glueball properties at finite temperature in SU(3) anisotropic lattice
  %QCD,''
  Phys.\ Rev.\  D {\bf 66}, 094506 (2002)
  [arXiv:hep-lat/0206020].
  %%CITATION = PHRVA,D66,094506;%%

%\cite{Meyer:2004jc}
\bibitem{Meyer:2004jc}
  H.~B.~Meyer and M.~J.~Teper,
  %``Glueball Regge trajectories and the pomeron: A Lattice study,''
  Phys.\ Lett.\  B {\bf 605}, 344 (2005)
  [arXiv:hep-ph/0409183].
  %%CITATION = PHLTA,B605,344;%%

%\cite{Nakamura:2010zzi}
\bibitem{Nakamura:2010zzi}
  KNakamura {\it et al.} [ Particle Data Group Collaboration ],
  %``Review of particle physics,''
  J.\ Phys.\ G {\bf G37}, 075021 (2010).

%\cite{Bali:1993fb}
\bibitem{Bali:1993fb}
  G.~S.~Bali, {\it et al.} % K.~Schilling, A.~Hulsebos, A.~C.~Irving, C.~Michael and P.~W.~Stephenson
                  [UKQCD Collaboration],
  %``A Comprehensive lattice study of SU(3) glueballs,''
  Phys.\ Lett.\  B {\bf 309}, 378 (1993);
  %[arXiv:hep-lat/9304012]

  % m_{0^{++}}=1.550\pm0.050
  %%CITATION = PHLTA,B309,378;%%

%\cite{Yuan:2009vs}
\bibitem{Yuan:2009vs}
  X.~H.~Yuan and L.~Tang,
  %``Fermion correction to the mass of the scalar glueball in QCD sum rule,''
  Commun.\ Theor.\ Phys.\  {\bf 54}, 495 (2010)
  [arXiv:0911.0806 [hep-ph]].
  %%CITATION = CTPMD,54,495;%%

%\cite{Bagan:1990sy}
\bibitem{Bagan:1990sy}
  E.~Bagan and T.~G.~Steele,
  %``MASS OF THE SCALAR GLUEBALL: HIGHER LOOP EFFECTS IN THE QCD SUM RULES,''
  Phys.\ Lett.\  B {\bf 243}, 413 (1990).
  %%CITATION = PHLTA,B243,413;%%

%\cite{Reinders:1981ww}
\bibitem{Reinders:1981ww}
  L.~J.~Reinders, S.~Yazaki and H.~R.~Rubinstein,
  %``L = 1 LIGHT QUARK MESONS IN QCD,''
  Nucl.\ Phys.\  B {\bf 196}, 125 (1982).
  %%CITATION = NUPHA,B196,125;%%

%\cite{Burcham:1995jg}
\bibitem{Burcham:1995jg}
  W.~E.~Burcham and M.~Jobes,
  %``Nuclear and particle physics,''
%\href{http://www.slac.stanford.edu/spires/find/hep/www?irn=3158713}{SPIRES entry}
{\it  Harlow, UK: Longman (1995) 752 p}

%\cite{ROSENFELD:1967zz}
\bibitem{ROSENFELD:1967zz}
  A.~H.~ROSENFELD {\it et al.},
  %``Data on Particles and Resonant States,''
  Rev.\ Mod.\ Phys.\  {\bf 39}, 1 (1967).
  %%CITATION = RMPHA,39,1;%%

%\cite{Anisovich:1996zj}
\bibitem{Anisovich:1996zj}
  A.~V.~Anisovich, V.~V.~Anisovich and A.~V.~Sarantsev,
  %``$0^{++}$-Glueball/$q \bar q$-State Mixing in the Mass Region near 1500
  %MeV,''
  Phys.\ Lett.\  B {\bf 395}, 123 (1997)
  [arXiv:hep-ph/9611333].
  %%CITATION = PHLTA,B395,123;%%

%\cite{Close:2000yk}
\bibitem{Close:2000yk}
  F.~E.~Close and A.~Kirk,
  %``The mixing of the f0(1370), f0(1500) and f0(1710) and the search for  the
  %scalar glueball,''
  Phys.\ Lett.\  B {\bf 483}, 345 (2000)
  [arXiv:hep-ph/0004241].
  %%CITATION = PHLTA,B483,345;%%

%\cite{Close:2001ga}
\bibitem{Close:2001ga}
  F.~E.~Close and A.~Kirk,
  %``Scalar glueball q anti-q mixing above 1-GeV and implications for lattice
  %QCD,''
  Eur.\ Phys.\ J.\  C {\bf 21}, 531 (2001)
  [arXiv:hep-ph/0103173].
  %%CITATION = EPHJA,C21,531;%%

%\cite{Giacosa:2005zt}
\bibitem{Giacosa:2005zt}
  F.~Giacosa, T.~Gutsche, V.~E.~Lyubovitskij and A.~Faessler,
  %``Scalar nonet quarkonia and the scalar glueball: mixing and decays in an
  %effective chiral approach,''
  Phys.\ Rev.\  D {\bf 72}, 094006 (2005)
  [arXiv:hep-ph/0509247].
  %%CITATION = PHRVA,D72,094006;%%

%\cite{He:2006ij}
\bibitem{He:2006ij}
  X.~G.~He, X.~Q.~Li, X.~Liu and X.~Q.~Zeng,
  %``X(1812) in quarkonia-glueball-hybrid mixing scheme,''
  Phys.\ Rev.\  D {\bf 73}, 114026 (2006)
  [arXiv:hep-ph/0604141].
  %%CITATION = PHRVA,D73,114026;%%

%\cite{Shifman:2010zza}
\bibitem{Shifman:2010zza}
  M.~Shifman,
  %``QCD sum rules: Bridging the gap between short and large distances,''
  Nucl.\ Phys.\ Proc.\ Suppl.\  {\bf 207-208}, 298 (2010)
  [arXiv:1101.1122 [hep-ph]].
  %%CITATION = NUPHZ,207-208,298;%%

%\cite{Narison:1997nw}
\bibitem{Narison:1997nw}
  S.~Narison,
  %``Masses, decays and mixings of gluonia in QCD: A summary,''
  Nucl.\ Phys.\ Proc.\ Suppl.\  {\bf 64}, 210 (1998)
  [arXiv:hep-ph/9710281].
  %%CITATION = NUPHZ,64,210;%%

%\cite{Narison:1996fm}
\bibitem{Narison:1996fm}
  S.~Narison,
  %``Masses, decays and mixings of gluonia in QCD,''
  Nucl.\ Phys.\  B {\bf 509}, 312 (1998)
  [arXiv:hep-ph/9612457].
  %%CITATION = NUPHA,B509,312;%%

%\cite{Harnett:2008cw}
\bibitem{Harnett:2008cw}
  D.~Harnett, R.~T.~Kleiv, K.~Moats and T.~G.~Steele,
  %``Near-Maximal Mixing of Scalar Gluonium and Quark Mesons: A Gaussian
  %Sum-Rule Analysis,''
  Nucl.\ Phys.\  A {\bf 850}, 110 (2011)
  [arXiv:0804.2195 [hep-ph]].
  %%CITATION = NUPHA,A850,110;%%

%\cite{Huang:1998wj}
\bibitem{Huang:1998wj}
  T.~Huang, H.~-Y.~Jin, A.~-L.~Zhang,
  %``Determination of the scalar glueball mass in QCD sum rules,''
  Phys.\ Rev.\  {\bf D59}, 034026 (1999).
  [hep-ph/9807391].

%\cite{Du:2004ki}
\bibitem{Du:2004ki}
  D.~S.~Du, J.~W.~Li and M.~Z.~Yang,
  %``Mass and decay constant of I = 1/2 scalar meson in QCD sum rule,''
  Phys.\ Lett.\  B {\bf 619}, 105 (2005)
  [arXiv:hep-ph/0409302].
  %%CITATION = PHLTA,B619,105;%%

%\cite{Pascual:1984zb}
\bibitem{Pascual:1984zb}
  P.~Pascual and R.~Tarrach,
  %``QCD: RENORMALIZATION FOR THE PRACTITIONER,''
  Lect.\ Notes Phys.\  {\bf 194}, 1 (1984).
  %%CITATION = LNPHA,194,1;%%

%\cite{Kataev:1981gr}
\bibitem{Kataev:1981gr}
  A.~L.~Kataev, N.~V.~Krasnikov and A.~A.~Pivovarov,
  %``TWO LOOP CALCULATIONS FOR THE PROPAGATORS OF GLUONIC CURRENTS,''
  Nucl.\ Phys.\  B {\bf 198}, 508 (1982)
  [Erratum-ibid.\  B {\bf 490}, 505 (1997)]
  [arXiv:hep-ph/9612326].
  %%CITATION = NUPHA,B198,508;%%

%\cite{Ablikim:2004wn}
\bibitem{Ablikim:2004wn}
  M.~Ablikim {\it et al.}  [BES Collaboration],
  %``Resonances in J / psi ---> phi pi+ pi- and phi K+ K-,''
  Phys.\ Lett.\  B {\bf 607}, 243 (2005)
  [arXiv:hep-ex/0411001].
  %%CITATION = PHLTA,B607,243;%%

%\cite{Ablikim:2006dw}
\bibitem{Ablikim:2006dw}
  M.~Ablikim {\it et al.}  [BES Collaboration],
  %``Observation of a near-threshold enhancement in the omega phi mass spectrum
  %from the doubly OZI suppressed decay J / psi --> gamma omega phi,''
  Phys.\ Rev.\ Lett.\  {\bf 96}, 162002 (2006)
  [arXiv:hep-ex/0602031].
  %%CITATION = PRLTA,96,162002;%%

%\cite{He:2006tw}
\bibitem{He:2006tw}
  X.~G.~He, X.~Q.~Li, X.~Liu and X.~Q.~Zeng,
  %``New members in the 0+ (0++) family,''
  Phys.\ Rev.\  D {\bf 73}, 051502 (2006)
  [arXiv:hep-ph/0602075].
  %%CITATION = PHRVA,D73,051502;%%

%\cite{Colangelo:2000dp}
\bibitem{Colangelo:2000dp}
  P.~Colangelo and A.~Khodjamirian,
  %``QCD sum rules, a modern perspective,''
  arXiv:hep-ph/0010175.
  %%CITATION = HEP-PH/0010175;%%

%\cite{Zhang:2003mr}
\bibitem{Zhang:2003mr}
  A.~l.~Zhang and T.~G.~Steele,
  %``Instanton and higher-loop perturbative contributions to the QCD  sum-rule
  %analysis of pseudoscalar gluonium,''
  Nucl.\ Phys.\  A {\bf 728}, 165 (2003)
  [arXiv:hep-ph/0304208].
  %%CITATION = NUPHA,A728,165;%%

%\cite{Narison:1984hu}
\bibitem{Narison:1984hu}
  S.~Narison,
  %``SPECTRAL FUNCTION SUM RULES FOR GLUONIC CURRENTS,''
  Z.\ Phys.\  C {\bf 26}, 209 (1984).
  %%CITATION = ZEPYA,C26,209;%%

%\cite{Narison:1988ts}
\bibitem{Narison:1988ts}
  S.~Narison and G.~Veneziano,
  %``QCD TESTS OF G (1.6) = GLUEBALL,''
  Int.\ J.\ Mod.\ Phys.\  A {\bf 4}, 2751 (1989).
  %%CITATION = IMPAE,A4,2751;%%

%\cite{Mathieu:2008me}
\bibitem{Mathieu:2008me}
  V.~Mathieu, N.~Kochelev, V.~Vento,
  %``The Physics of Glueballs,''
  Int.\ J.\ Mod.\ Phys.\  {\bf E18}, 1-49 (2009).
  [arXiv:0810.4453 [hep-ph]].

%\cite{Shuiguo:2010ak}
\bibitem{Shuiguo:2010ak}
  W.~Shuiguo, Z.~Zhenyu, L.~Jueping,
  %``$0^{++}$ scalar glueball in finite-width Gaussian sum rules,''
  Phys.\ Rev.\  {\bf D82}, 016003 (2010).
  [arXiv:1007.2465 [hep-ph]].

%\cite{Narison:1984bv}
\bibitem{Narison:1984bv}
  S.~Narison, N.~Pak and N.~Paver,
  %``MESON - GLUONIUM MIXING FROM QCD SUM RULES,''
  Phys.\ Lett.\  B {\bf 147}, 162 (1984).
  %%CITATION = PHLTA,B147,162;%%

%\cite{Mennessier:1985mz}
\bibitem{Mennessier:1985mz}
  G.~Mennessier, S.~Narison and N.~Paver,
  %``GLUONIUM AND THE 0++ SPECTRUM,''
  Phys.\ Lett.\  B {\bf 158}, 153 (1985).
  %%CITATION = PHLTA,B158,153;%%

\end{thebibliography}
\end{document}